\documentclass[preprint,aps]{revtex4}

\usepackage{graphicx,epsfig}% Include figure files
\usepackage{float}
\usepackage{dcolumn}% Align table columns on decimal point
\usepackage{bm}% bold math

\begin{document}

%\preprint{APS/123-QED}

\title{Attempt frequency of magnetization in nanomagnets with thin-film geometry}
\author{Hong-Ju Suh$^1$, Changehoon Heo$^1$, Chun-Yeol You$^2$, Woojin Kim$^3$, Taek-Dong Lee$^3$, and Kyung-Jin Lee$^1$$^\dagger$}
\affiliation{$^1$Department of Materials Science and Engineering, Korea University, Seoul 136-701, Korea \\
$^2$Department of Physics, Inha University, Incheon 402-751, Korea \\
$^3$Department of Materials Science and Engineering, KAIST, Daejon 305-701, Korea}

\date{\today}% It is always \today, today,

\begin{abstract}
Solving the stochastic Landau-Lifshitz-Gilbert equation numerically,
we investigate the effect of the potential landscape on the attempt
frequency of magnetization in nanomagnets with the thin-film
geometry. Numerical estimates of the attempt frequency are analyzed
in comparison with theoretical predictions from the Fokker-Planck
equation for the N\'{e}el-Brown model. It is found that for a
nanomagnet with the thin-film geometry, theoretically predicted
values for the universal case are in excellent agreement with
numerical estimates.
\end{abstract}


\maketitle

%%%%%%%%%%%%%%%%%%%%%%%%%%%%%%%%%%%%%%%%%%%%%%%%%%%%%%%%%%%%%%%%%%%%%%%%%%%%%%
\section{Introduction}
Much effort has been expended in fabricating deep sub-micron
patterned magnets with the thin-film geometry. From the scientific
point of view, such a small magnet is a good model system to study
basic magnetism via a direct comparison with an idealized
theoretical prediction. From the application point of view, a steady
progress of the patterning technology for fabricating a smaller cell
has led to magnetic devices such as spin-valve read sensors for the
hard disk drive and magnetic random access memories (MRAMs)
utilizing the spin-transfer torque (STT)~\cite{Slon, BergerPRB} into
a higher density, i.e. a smaller magnetic volume.

Thermal agitation of a magnetization becomes more and more important
as the magnetic volume of a unit cell decreases. In the spin-valve
read sensor, the so-called "mag-noise" is a manifestation of the
thermally excited ferromagnetic resonance in the sensor
stack~\cite{Smith, ZhuJAP}. In the STT-MRAM, thermal agitation
hinders a continuous miniaturization of the device because it can
cause spontaneous changes of magnetization direction from one stable
state to another.

Thermal relaxation time is a statistical time-scale for which a
magnetization escapes from an initial local minimum state over an
energy barrier. The thermal relaxation time ${\tau}$ of a
magnetization is described by the N\'{e}el-Brown model~\cite{Neel,
BrownPR} in the high energy barrier asymptote,
${\tau}={f}_{0}^{-1}\exp [{{U_B} / {k_B T}} ]$ where $f_0$ is the
attempt frequency, $U_B$ is the energy barrier measuring the
difference between a local minimum and a saddle point, $k_B$ is the
Boltzmann constant, and  $T$ is the temperature in Kelvin.

Experimental studies on the thermal relaxation of magnetization
generally assume a constant attempt frequency~\cite{WernsdorferPRL,
Rizzo, Woods}. However, Brown showed theoretically that the attempt
frequency is not constant but depends on many parameters such as the
damping constant and the magnetic properties~\cite{BrownPR}.
Followed by Brown's initial work~\cite{BrownPR}, theoretical
formulae of the attempt frequency for different potential symmetry
were proposed~\cite{BrownIEEE, Kilk, CoffeyPRB, CoffeyACP,
Dejardin,CoffeyPSS,Garanin,Kalmykov}.

Accurate theoretical formulae of the attempt frequency are necessary
for modeling experiments and predicting quantitatively the
superparamagnetic limit for device applications. However, it is not
easy to experimentally verify the theoretical formulae because i) an
experimentally measurable quantity such as the switching field is
mostly governed by the energy barrier, not by the attempt frequency,
and ii) the damping constant, a key factor affecting the attempt
frequency, of a small magnet is not definite in
general~\cite{CoffeyPRL, Andrade}.

In this work, by means of a numerical study based on the stochastic
Landau-Lifshitz-Gilbert (LLG) equation~\cite{BrownPR, Chantrell}, we
investigate the validity of the proposed theoretical formulae. It is
found that for a nanomagnet with the thin-film geometry,
theoretically predicted values for the universal case are in
excellent agreement with numerical estimates whereas theoretical
values for the intermediate-to-high damping limit and the very low
damping limit fail to reproduce numerical ones in practically
meaningful ranges of the damping constant.

This paper is organized as follows. After introducing the proposed
theoretical formulae (Sec. II) and numerical model used in this work
(Sec. III), we show in Sec. IV the effect of the shape anisotropy,
i.e. potential landscape, on the attempt frequency for various
damping constants and discuss about validity of the theoretical
formulae by comparing with numerical estimates. In Sec V, we
summarize this work.

%%%%%%%%%%%%%%%%%%%%%%%%%%%%%%%%%%%%%%%%%%%%%%%%%%%%%%%%%%%%%%%%%%%%%%%%%%%%%%%%%%%%%
\begin{figure}[ttbp]
\begin{center}
\psfig{file=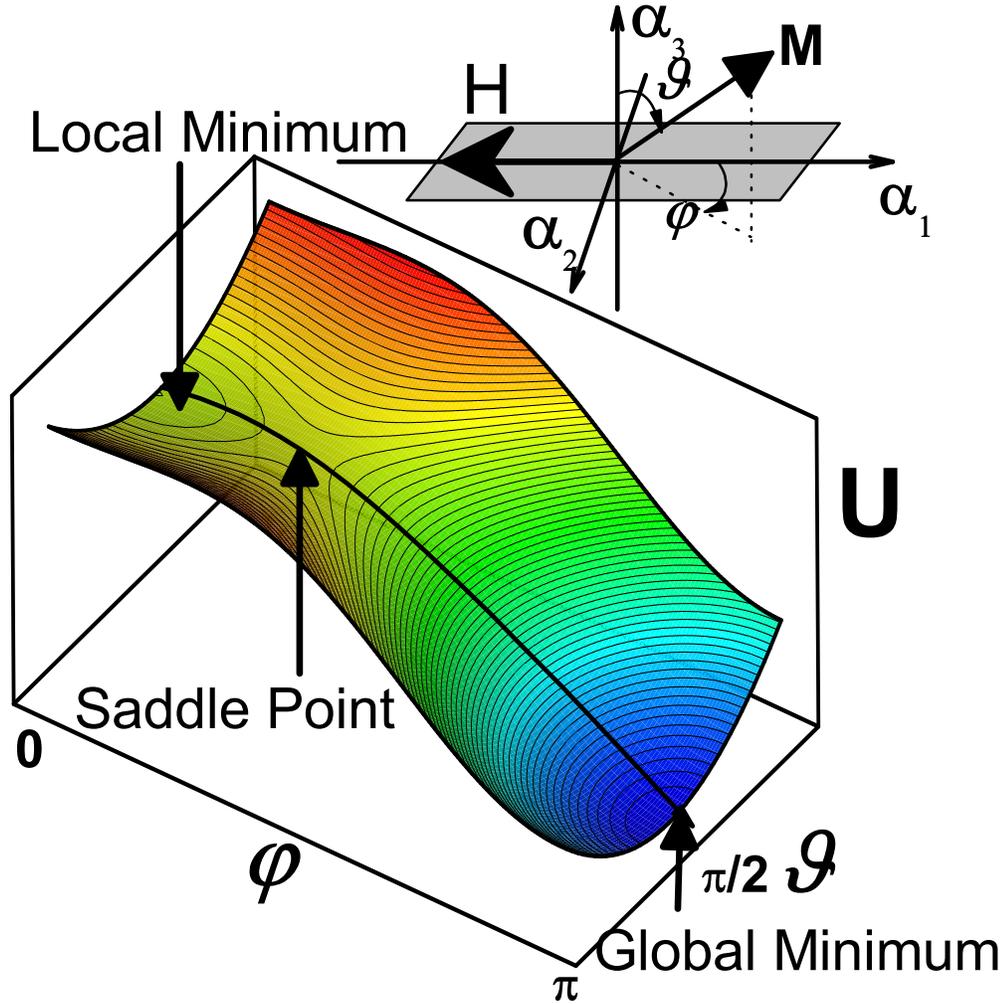,width=\columnwidth} \caption{\label{fg1} (Color
online) Magnetic potential surface of a single domain particle with
non-axial symmetry. $\alpha_{1}$ is the magnetic easy axis and the
external field $H$ is applied along the easy axis. The magnetic
energy $U$ in the Eq. (1) has two equivalent saddle points and two
minima points; local minimum and global minimum. In thermally
activated switching, the magnetization changes from local minimum to
global minimum, passing through saddle point.}
\end{center}
\end{figure}
%%%%%%%%%%%%%%%%%%%%%%%%%%%%%%%%%%%%%%%%%%%%%%%%%%%%%%%%%%%%%%%%%%%%%%%%%%%%%%%%%%%%%%%%

\section{Theoretical formulae of the attempt frequency}
The magnetic potential $U$ of a single-domain particle with uniaxial
symmetry in the presence of a static external longitudinal field $H$
is given by
%%%%%%%%%%%%%%%%%%%%%%%%%%%%%%%%%%%%%%%%%%%%%%%%%%%%%%%%%%%%%%%%%%%%%%%%%%%%%%%%%%%%%%%%%%%%

\begin{eqnarray}\label{potential}
\lefteqn{ U=K_{U}(1-\alpha_{1}^{2})-M_{S}H\alpha_{1}+ {1 \over 2}( N_{1}\alpha_{1}^{2}+N_{2}\alpha_{2}^{2}+N_{3}\alpha_{3}^{2})M_{s}^{2} {} }
\nonumber\\& & {}
 =K_{U}(1-\sin^{2}\vartheta\cos^{2}\varphi)-M_{S}H\sin\vartheta\cos\varphi {}
\nonumber\\& & {}
  +{1 \over 2} (N_{1}\sin^{2}\vartheta\cos^{2}\varphi +N_{2}\sin^{2}\vartheta\sin^{2}\varphi+N_{3}\cos^{2}\vartheta)M_{s}^{2}, {}
\end{eqnarray}
%%%%%%%%%%%%%%%%%%%%%%%%%%%%%%%%%%%%%%%%%%%%%%%%%%%%%%%%%%%%%%%%%%%%%%%%%%%%%%%%%%%%%%%%%%%%%%
where $K_{U}$ is the uniaxial anisotropy, $M_S$ is the saturation
magnetization, $H$ is the external field applied along the magnetic
easy axis, ${\vartheta}$ and ${\varphi}$ are the polar angle of
magnetization vector and the azimuthal angle of magnetization
vector, respectively, ${\alpha_{i}}$ (i = 1, 2, 3) is the direction
cosines of magnetization vector, and $N_{i}$ is the demagnetization
factor along ${\alpha_{i}}$ axis.

 In an axially symmetric potential ($U(\vartheta,\varphi) = U( \vartheta )$), Brown~\cite{BrownPR} showed the attempt frequency is not a constant, but a complex function as
%%%%%%%%%%%%%%%%%%%%%%%%%%%%%%%%%%%%%%%%%%%%%%%%%%%%%%%%%%%%%%%%%%%%%%%%%%%%%%%%%%%%%
\begin{equation}\label{Brown1}
{f_0={\gamma \alpha \over 1+\alpha^2} \sqrt{H_K^3 M_S V \over 2 \pi
k_B T} (1-h^2 )(1+h). }
\end{equation}
%%%%%%%%%%%%%%%%%%%%%%%%%%%%%%%%%%%%%%%%%%%%%%%%%%%%%%%%%%%%%%%%%%%%%%%%%%%%%%%%%%%%%
where ${\gamma}$ is the gyromagnetic ratio, ${\alpha}$ is the
damping constant, $H_K$ is the effective anisotropy field, $V$ is the magnetic volume, $h$ is
$H/H_K$, and $H$ is the external field applied along the magnetic
easy axis.

Dependence of the attempt frequency on the damping constant for a
non-axially symmetric potential was first theoretically predicted
for two limiting cases; i) intermediate-to-high damping (IHD)
case~\cite{BrownIEEE} , and ii) very low damping (VLD)
case~\cite{Kilk, CoffeyPRB}. Later, the universal theoretical
equation~\cite{CoffeyACP, Dejardin,CoffeyPSS,Garanin,Kalmykov} which
is valid for all values of the damping constant was derived by
extending the Meshkov-Mel'ikov depopulation factor to the magnetic
case~~\cite{Meshkov}. In this work, numerical estimates of the
attempt frequencies are compared with the theories for the two
limiting cases, and the universal one.

In the IHD limit, the attempt frequency for a non-axially symmetric
potential is given as~\cite{BrownIEEE},
%%%%%%%%%%%%%%%%%%%%%%%%%%%%%%%%%%%%%%%%%%%%%%%%%%%%%%%%%%%%%%%%%%%%%%%%%%%%%%%%%%%%%
\begin{equation}\label{Brown2}
%\begin{eqnarray}\label{Brown2}
f_{0}^{IHD}={\gamma \alpha \sqrt{c_{m1} c_{m2}} \over 4 \pi M_S (1+\alpha^2
)} {(c_2^\prime -c_1 )+\sqrt{(c_1+c_2^\prime )^2 + 4c_1 c_2^\prime /
\alpha^2} \over \sqrt{c_1 c_2^\prime} },
%\end{eqnarray}
\end{equation}
%%%%%%%%%%%%%%%%%%%%%%%%%%%%%%%%%%%%%%%%%%%%%%%%%%%%%%%%%%%%%%%%%%%%%%%%%%%%%%%%%%%%%
where $c_{m1}$ and $c_{m2}$ are the coefficients in the expansion of
magnetic potential $U$ about a local energy minimum for the initial
magnetic state, $U=U_m +1/2(c_{m1} \alpha_1^2 +c_{m2} \alpha_2^2
)+\cdots$ ; $c_1$ and $c_2^\prime$ are the coefficients in the
expansion about the saddle point, $U=U_S +1/2(c_1 \alpha_1^2 -
c_2^\prime \alpha_2^2 )+\cdots$, respectively.

Klik and Gunther~\cite{Kilk}, and Coffey~\cite{CoffeyPRB} derived a
theoretical formalism of the attempt frequency for a non-axially
symmetric potential in the VLD limit as,
%%%%%%%%%%%%%%%%%%%%%%%%%%%%%%%%%%%%%%%%%%%%%%%%%%%%%%%%%%%%%%%%%%%%%%%%%%%%%%%%%%%%%
\begin{equation}\label{Kilk}
{f_{0}^{VLD}={{\gamma \alpha \sqrt{c_{m1} c_{m2}} \over 2 \pi M_S (1+\alpha^2
) }}S, }
\end{equation}
%%%%%%%%%%%%%%%%%%%%%%%%%%%%%%%%%%%%%%%%%%%%%%%%%%%%%%%%%%%%%%%%%%%%%%%%%%%%%%%%%%%%%
where S is the dimensionless action variable at the saddle point potential $U_S$ defined as
%%%%%%%%%%%%%%%%%%%%%%%%%%%%%%%%%%%%%%%%%%%%%%%%%%%%%%%%%%%%%%%%%%%%%%%%%%%%%%%%%%%%%
\begin{equation}\label{Saddle}
{S=\frac{V}{k_{B}T}  \oint_{U(\vartheta, \varphi)=U_{S}}  \Big[(1-\cos^{2}\vartheta) \frac{\partial}{\partial\cos\vartheta} U (\vartheta, \varphi) d \varphi
- { 1 \over ( 1- \cos^{2}\vartheta)}
\frac{\partial}{\partial\varphi} U (\vartheta,\varphi) d \cos \vartheta \Big].}
\end{equation}
%%%%%%%%%%%%%%%%%%%%%%%%%%%%%%%%%%%%%%%%%%%%%%%%%%%%%%%%%%%%%%%%%%%%%%%%%%%%%%%%%%%%%%

For the universal case, the attempt frequency is given
by~\cite{CoffeyACP, Dejardin,CoffeyPSS,Garanin,Kalmykov},
%%%%%%%%%%%%%%%%%%%%%%%%%%%%%%%%%%%%%%%%%%%%%%%%%%%%%%%%%%%%%%%%%%%%%%%%%%%%%%%%%%%%%%
\begin{equation}\label{cross}
{f_{0}=A(\alpha S)f_{0}^{IHD},}
\end{equation}
%%%%%%%%%%%%%%%%%%%%%%%%%%%%%%%%%%%%%%%%%%%%%%%%%%%%%%%%%%%%%%%%%%%%%%%%%%%%%%%%%%%%%
where S is given by Eq.~(\ref{Saddle}).

$A(\alpha S)$ is a factor which interpolates between the VLD and IHD
limits, and given by
%%%%%%%%%%%%%%%%%%%%%%%%%%%%%%%%%%%%%%%%%%%%%%%%%%%%%%%%%%%%%%%%%%%%%%%%%%%%%%%%%%%%%%
\begin{equation}\label{A}
{A(\alpha S)=\exp \Big[{1 \over \pi} \int_{0}^{\infty}\frac{ \ln
[1-\exp \{ -\alpha S(\lambda^{2} + 1/4) \}]} {\lambda^{2} + 1/4} d
\lambda \Big].}
\end{equation}
%%%%%%%%%%%%%%%%%%%%%%%%%%%%%%%%%%%%%%%%%%%%%%%%%%%%%%%%%%%%%%%%%%%%%%%%%%%%%%%%%%%%%%%

%%%%%%%%%%%%%%%%%%%%%%%%%%%%%%%%%%%%%%%%%%%%%%%%%%%%%%%%%%%%%%%%%%%%%%%%%%%%%%%%%%%%%
\begin{figure}[ttbp]
\begin{center}
\psfig{file=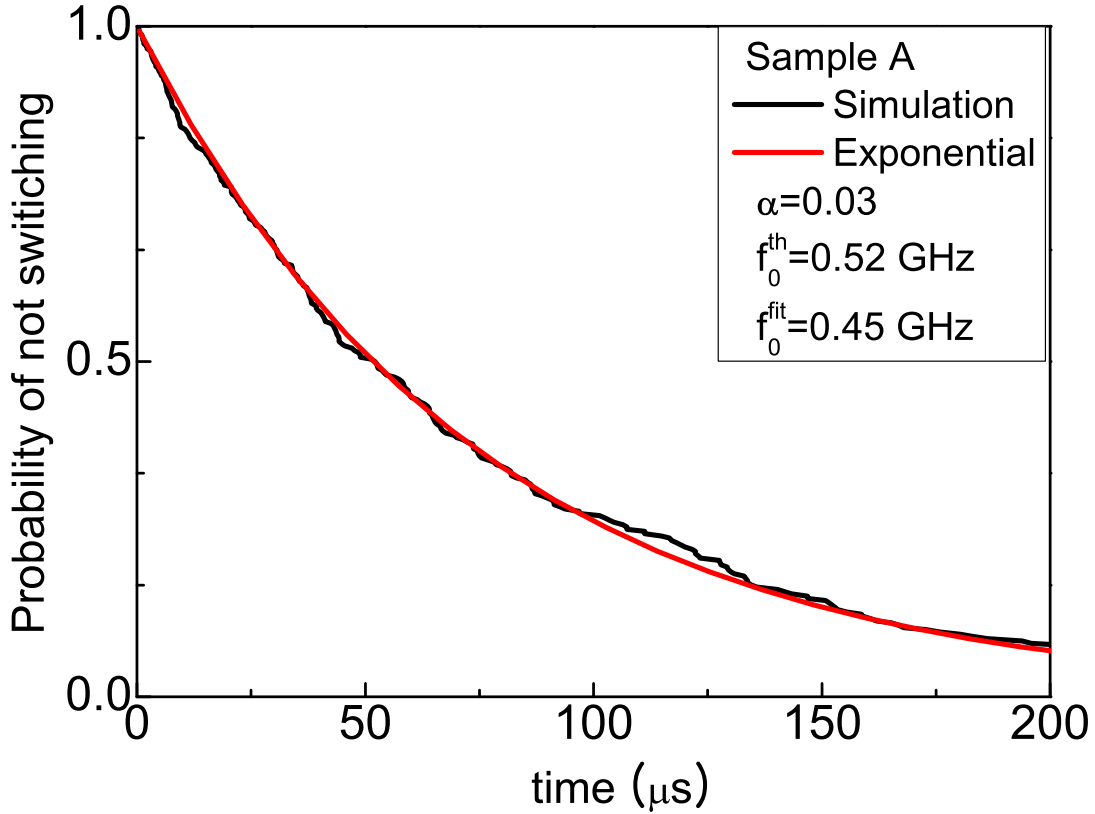,width=\columnwidth} \caption{\label{fg2} (Color
online) Probability of not switching ($1 -P_{SW}$) versus
magnetization switching time at $\alpha$=0.03 in the sample A ($(l
\times w \times d=21 \times 20 \times 20 nm^3$).}
\end{center}
\end{figure}
%%%%%%%%%%%%%%%%%%%%%%%%%%%%%%%%%%%%%%%%%%%%%%%%%%%%%%%%%%%%%%%%%%%%%%%%%%%%%%%%%%%%%

\section{Numerical Model}
We performed macrospin calculations by means of
the stochastic LLG equation,

%%%%%%%%%%%%%%%%%%%%%%%%%%%%%%%%%%%%%%%%%%%%%%%%%%%%%%%%%%%%%%%%%%%%%%%%%%%%%%%%%%%%%
\begin{equation}\label{LLG}
{\partial \mathbf{M} \over \partial t}=-\gamma \mathbf{M} \times
\mathbf{H_{eff}} + {\alpha \over M_S} {\partial \mathbf{M} \over
\partial t}
\end{equation}
%%%%%%%%%%%%%%%%%%%%%%%%%%%%%%%%%%%%%%%%%%%%%%%%%%%%%%%%%%%%%%%%%%%%%%%%%%%%%%%%%%%%%
where $\textbf{M}$ is the magnetization vector, and
$\textbf{H}_\textbf{eff}$ is the effective magnetic field including
the external, the magnetostatic, the thermal fluctuation. To
estimate the thermal relaxation time ${\tau}$, we used a macrospin
model with ${U_B}/{k_B T}$ of about $10$ because of excessive
computation time. Probability of switching $P_{SW}$ of
thermally-activated switching was estimated by counting the number
of successful switching out of $500$ switching events. The attempt
frequency is obtained by fitting numerical results of $P_{SW}$ as a
function of the time using the Arrhenius-N\'{e}el decay of the
probability of switching, $P_{SW}=1-\exp[-f_0 t \exp(-U_B /k_B T)]$
as shown in Fig. ~\ref{fg2}.

\section{Effect of shape anisotropy on the attempt frequency}
We have calculated attempt frequencies of various sized nanomagnets;
sample A $(l \times w \times d=21 \times 20 \times 20 nm^3$), sample
B ($25 \times 21 \times 16 nm^3$), sample C ($40 \times 30 \times 7
nm^3$), and sample D ($100 \times 28 \times 3 nm^3$) where $l$
(length), $w$ (width), and $d$ (thickness) are the sample dimensions
along $x$, $y$, and $z$-axis, respectively and thus the $x$-axis is
the easy axis. For all four samples, constant values of volume $V (=
8400 nm^3 )$, effective in-plane anisotropy $H_K (= 875.4 Oe)$, and
external field H (= -540 Oe) were used to exclude their effects on
the attempt frequency. The thermal stability factor $U_B /k_B T$ was
10.425, a good number for the high energy barrier approximation. The
magnetic potential $U$ of the sample A is axially symmetric since
$w=d$, whereas $U$ of other samples are non-axially symmetric since
$w \neq d$.

\begin{figure}[ttbp]
\begin{center}
\psfig{file=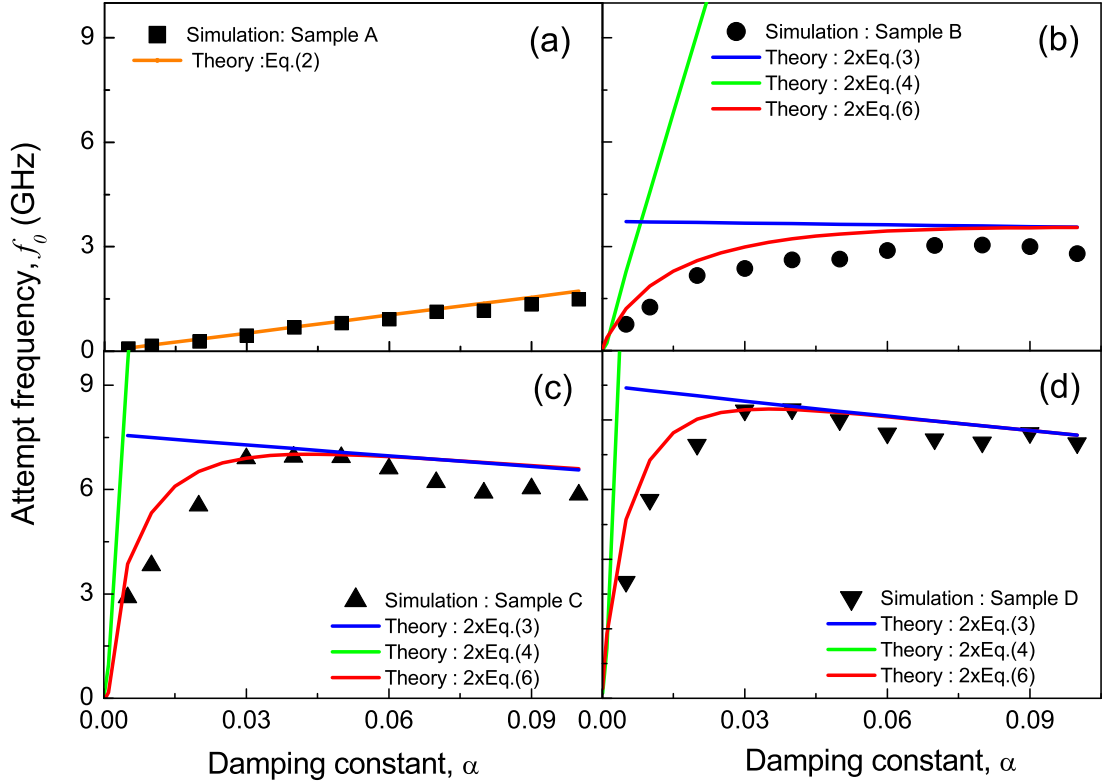,width=\columnwidth} \caption{\label{fg3} (Color
online)(a), (b), (c), and (d) show the attempt frequency as a
function of the damping constant $\alpha$ for the sample A, B, C,
and D, respectively. Solid lines are theoretically predicted values
and symbols are numerical results. Dimensions of samples and
parameters are given in the text ($M_S=800 emu/cm^3)$.}
\end{center}
\end{figure}

The numerical results of the attempt frequency for the nanomagnet
with an axially symmetric potential (sample A) are shown in Fig.
~\ref{fg3}(a). To our knowledge, the Eq.~(\ref{Brown1}) was tested
once by adopting the same way used in this work~\cite{Boerner}, and
it was reported that the theoretical value of attempt frequency is
different from the numerically estimated value by an order of
magnitude. This inconsistency may have prevented further numerical
studies on the attempt frequency. However, we found excellent
agreement between the Eq.~(\ref{Brown1}) and numerically estimated
values (see Fig. 3(a)). The difference between the result in the
Ref. ~\cite{Boerner} and ours originates from the sign of $h$ in the
Eq.~(\ref{Brown1}). The $h$ should be negative since the
magnetization is initially in a shallower local energy minimum,
whereas it was assumed to be positive in the Ref.~\cite{Boerner}.
The excellent agreement verifies the validity of our numerical
approach to estimate the attempt frequency in this work.

%%%%%%%%%%%%%%%%%%%%%%%%%%%%%%%%%%%%%%%%%%%%%%%%%%%%%%%%%%%%%%%%%%%%%%%%%%%%%%%%%%%%%
%%%%%%%%%%%%%%%%%%%%%%%%%%%%%%%%%%%%%%%%%%%%%%%%%%%%%%%%%%%%%%%%%%%%%%%%%%%%%%%%%%%%%

Fig. ~\ref{fg3}(b), (c), and (d) show the dependence of the attempt
frequency on the damping constant for the sample B, C, and D,
respectively. Two features are worth mentioning. First, the attempt
frequency increases with increasing $w/d$. For instance, the attempt
frequency of sample D is an order of magnitude higher than that of
sample A in the wide range of damping constant (Fig.~\ref{fg3}).
Second, when the potential landscape is non-axially symmetric, there
are two regimes of the damping constant where the attempt frequency
shows an explicitly different dependence on the damping constant. At
low damping ($\alpha < 0.03$), the attempt frequency increases with
the damping constant whereas at high damping
($\alpha > 0.03$), it slightly decreases.

Considering the increase of attempt frequency with $w/d$, it should
be noted that both Eqs.~(\ref{Brown2}) and ~(\ref{Kilk}) contain
$\sqrt{c_{m1} c_{m2}}$ which is an averaged curvature of potential
at the local minimum. Other terms do not vary much with $w/d$. The
$c_{m1}$ is given by $H_K M_S (1+h )$ which is a constant for all
four samples since both effective in-plane anisotropy field $H_K$
and external field $H$ are assumed to be constants. The $c_{m2}$ is
given by $H_K M_S(1+h+{{ (N_{3}-N_{2})M_S } \over H_K } )$ where
$N_2$ ($N_3$) is the demagnetization factor along the in-plane hard
(out-of-plane hard) axis. The $c_{m2}$ significantly varies with the
sample shape since ${ (N_{3}-N_{2})M_S } \over H_K $ term is
dominant. Therefore, an important parameter to determine the attempt
frequency is the coefficient $c_{m2}$ which measures the curvature
of potential along the direction cosine $\alpha_2$ and is related to
the out-of-plane demagnetization effect. A larger curvature of
potential at a local minimum results in a higher attempt frequency.
It is because the magnetization moving away from a local minimum due
to a thermal random force experiences an instantaneous restoring
force proportional to the curvature. The curvature $c_{m2}$ becomes
smaller and smaller as the aspect ratio of sample $w/d$ approaches
the unity. Among the tested samples, the sample D provides the
largest $c_{m2}$ and thus, the highest attempt frequency.

In order to understand dependence of the attempt frequency on the
damping constant, we compare numerical results with the theoretical
formulae (Eq. (\ref{Brown2}) (IHD) , (\ref{Kilk}) (VLD) and
(\ref{cross}) (Universal)) from the Fokker-Planck equation for the
N\'{e}el-Brown model. In the whole range of damping constant , the
numerical results are in good agreements with the Eq.~(\ref{cross})
multiplied by factor 2. The equations were derived for the escape of
magnetization over only one shallower barrier assuming different
barrier height between in-plane clockwise switching and counter
clockwise one. In our case, the two energy barriers are identical
since no symmetry breaking exists, validating the multiplication by
factor 2.

The theoretical values obtained from the Eq. (\ref{Brown2}) (IHD)
partially coincide with the numerical results in high damping regime
($\alpha > 0.04$), whereas the Eq. (\ref{Kilk}) (VLD) predicts much
higher attempt frequencies than the numerical results in the tested
range of damping constant ($0.005 < \alpha < 0.1$).

In the VLD, the escape rate is evaluated from the energy loss per
cycle of a particle on the escape rate trajectory~\cite{CoffeyPSS,
Kalmykov}. The assumption made in deriving the Eq. (\ref{Kilk}),
replacing the energy loss per cycle of the almost periodic motion at
the barrier energy by the barrier height, is necessarily crude and
only applies when the damping constant is less than about 0.001. The
failure of Eq. (\ref{Kilk}) to estimate the attempt frequency is
also found in the Ref.~\cite{Dejardin} where comparisons among the
IHD escape rates, the VLD escape rates, the universal solution based
on the Meshkov-Mel'inkov depopulation factor, and the exact escape
rate based on the continued fraction solution of the Fokker-Planck
for the lowest eigenvalue were made. In the Ref.~\cite{Dejardin}, it
is shown that the VLD asymptote begins to fail for the damping
constant of the order of $10^{-2}$ even if the action on the escape
trajectory is evaluated exactly whereas the universal solution
provides a reasonably accurate approximation throughout the whole
range of damping.

Therefore, it is obvious that the universal escape rate
(Eq.~(\ref{cross})) provides an accurate description of the
behaviour of the exact escape rate provided that the barrier height
is sufficient to allow one to define an escape rate. Furthermore,
since the damping constant in a typical nanomagnet with the thin
film geometry is in the range between 0.005 to
0.03~\cite{Krivorotov, SankeyPRL, Sankey, Kubota} where the VLD and
the IHD approximations show evidently wrong predictions for the
attempt frequency, the Eq. (\ref{cross}) should be used to design
experiments and to interpret experimental results performed at
non-zero temperatures.

\section{Summary}
In a nanomagnet with the thin-film geometry, the demagnetization
energy along the magnetic hard axis is a a main factor affecting the
attempt frequency. Comparing numerical estimates of the attempt
frequency of magnetization with the theoretically predicted values,
we verify the validity of the theoretical formula of the attempt
frequency for the universal case. However, the theoretical formulae
in the low damping limit and the intermediate-high damping limit
fail to reproduce numerical values for the typical range of the
damping constant. Therefore, the attempt frequency obtained from the
theoretical equation for the universal case should be used to design
experiments and to interpret experimental results performed at
non-zero temperatures.

%%%%%%%%%%%%%%%%%%%%%%%%%%%%%%%%%%%%%%%%%%%%%%%%%%%%%%%%%%%%%%%%%%%%%%%%%%%%%%%%%%%%%
\section*{Acknowledgments}
Comments by Hyun-Woo Lee and Sug-Bong Choe are appreciated. This
work is supported by the Korea Science and Engineering Foundation
(KOSEF) through the Basic Research Program funded by the Ministry of
Science and Technology (No. R01-2007-000-20281-0) and Samsung
Electronics.
%%%%%%%%%%%%%%%%%%%%%%%%%%%%%%%%%%%%%%%%%%%%%%%%%%%%%%%%%%%%%%%%%%%%%%%%%%%%%%%%%%%%%

($\dagger$) Corresponding email: kj\_lee@korea.ac.kr

%($\dagger$) Corresponding email: kj\_lee@korea.ac.kr.
%%%%%%%%%%%%%%%%%%%%%%%%%%%%%%%%%%%%%%%%%%%%%%%%%%%%%%%%%%%%%%%%%%%%%%%%%%%%%%%%%%%%%
%\\newpage %Just because of unusual number of tables stacked at end
%\\bibliography{apssamp}% Produces the bibliography via BibTeX.

%%%%%%%%%%%%%%%%%%%%%%%%%%%%%%%%%%%%%%%%%%%%%%%%%%%%%%%%%%%%%%%%%%%%%%%%%%%%%%%%%%%%%

\end{document}